\documentclass[a4paper, 10pt, oneside, twocolumn, 1p, number, sort&compress ]{elsarticle} %preprint 

\newcommand{\hho}{H$_2$O}
\newcommand{\oo}{O$_2$}
\newcommand{\oom}{O$_2^{\mbox{-}}$}

\newcommand{\srosurf}{Sr$_3$Ru$_2$O$_7$(001)}
\newcommand{\crosurf}{Ca$_3$Ru$_2$O$_7$(001)}
\newcommand{\cro}{Ca$_3$Ru$_2$O$_7$}
\newcommand{\sro}{Sr$_3$Ru$_2$O$_7$}

\newcommand{\opt}{optB86b}

\newcommand{\ea}{E_{\mathrm{ads}}}
\newcommand{\oha}{OH$_{\mathrm{ads}}$}
\newcommand{\osurf}{O$_{\mathrm{surf}}$}
\newcommand{\degree}{$^{\circ}$}
\newcommand{\stmpar}[3]{$V_{s} = {#1}\,\mathrm{V}, I_{t} = {#2}\,\mathrm{nA}, T = {#3}\,\mathrm{K}$}

\title{Adsorption of CO on the \crosurf~surface}
\author[iap,cms]{Wernfried Mayr-Schm\"olzer}
\ead{wms@cms.tuwien.ac.at}
\author[iap]{Daniel Halwidl}
\ead{halwidl@iap.tuwien.ac.at}
\author[iap,cms]{Florian Mittendorfer\corref{cor1}}
\ead{fmi@cms.tuwien.ac.at}
\author[iap]{Michael Schmid}
\ead{schmid@iap.tuwien.ac.at}
\author[iap]{Ulrike Diebold}
\ead{diebold@iap.tuwien.ac.at}
\author[iap,cms]{Josef Redinger}
\ead{jr@cms.tuwien.ac.at}
\cortext[cor1]{Corresponding author}
\address[iap]{Institute of Applied Physics, TU Wien, Wiedner Hauptstr. 8-10/134, 1040 Vienna, Austria}
\address[cms]{Center for Computational Materials Science, TU Wien, Wiedner Hauptstr. 8-10/134, 1040 Vienna, Austria}

\begin{document}

\begin{abstract}

The adsorption of CO molecules at the \crosurf\ surface was studied using low-temperature scanning tunneling microscopy (STM) and density functional theory (DFT). \cro\ can be easily cleaved along the (001) plane, yielding a smooth, CaO-terminated surface. 
The STM shows a characteristic pattern with alternating dark and bright stripes, resulting from the tilting of the RuO$_6$ octahedra.
%with a characteristic alternating wide-narrow channel structure that results from the tilting of the RuO$_6$ lattice octahedra. 
 At 78\,K, CO adsorbs at an apical surface O at the channel edge with a predicted binding energy of $\ea = -0.85$\,eV. After annealing at room temperature, the CO forms a strong bond ($\ea= -2.04$\,eV) with the apical O and the resulting carboxylate takes the place of the former surface O. This carboxylate can be decomposed by scanning the surface with a high sample bias voltage of +2.7\,V, restoring the original surface.
\end{abstract}


\maketitle

\section{Introduction}
The surface chemistry of complex ternary transition metal oxides has emerged as an important research topic, since these materials are increasingly used in electrocatalysis and solid oxide fuel cells.  A deeper understanding of surface properties on an atomic level is highly desirable. Surface-science investigations can deliver such insights, provided that appropriate model systems---single crystalline surfaces with a well-defined and known surface structure---are used. This is difficult for ternary materials, as the usual surface preparation method of sputtering and annealing affects the surface stoichiometry, which results in various, and often complex, surface reconstructions.  An elegant way to prepare simple (1$\times$1) terminated surfaces is cleaving appropriate layered materials. Here the Ruddelsden-Popper series A$_{n+1}$Ru$_{n}$O$_{3n+1}$ (A=Sr, Ca) has proven to be an ideal system, as high-quality crystals can be synthesized \cite{Mao2000} and easily cleaved in ultrahigh vacuum (UHV) \cite{Stoeger2014surf}.  The original impetus to grow such crystals stems from their interesting physical properties including superconductivity, ferroelectricity, or magnetoresistance, which can be tuned by external parameters such as doping, fields, pressure, or temperature. Recent research \cite{Stoeger2014co,Stoeger2014surf,Halwidl2015, Halwidl2017,Halwidl2018} has shown that they are also ideal model systems for investigating surface reactivity at the atomic scale.  

\begin{figure}[htbp]
\begin{center}
\includegraphics[trim= 0cm 0cm 0cm 0cm, clip, width=0.75\textwidth] {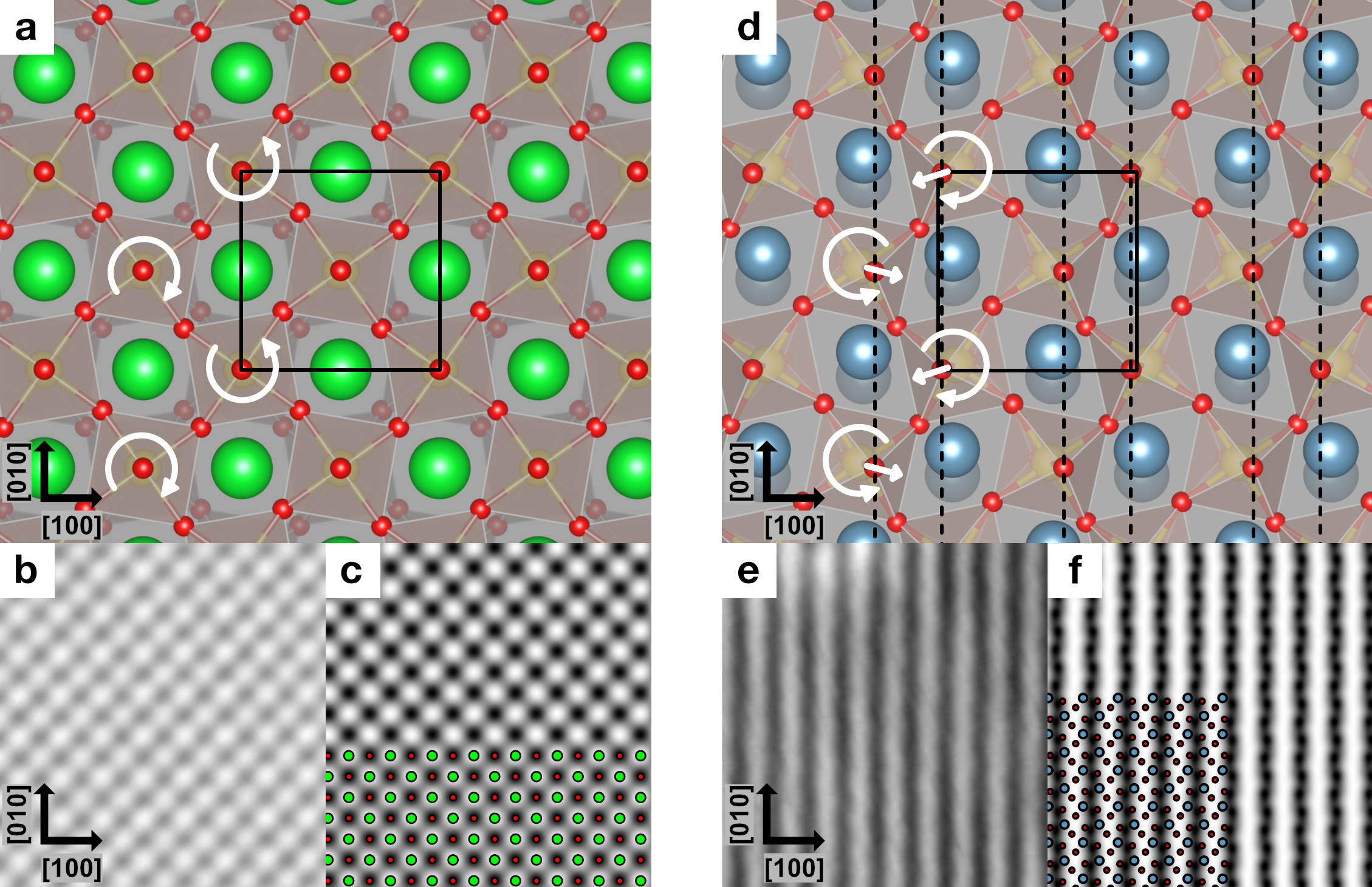}
\caption{\textbf{Surface structure of \sro~and \cro.} Ca \--{} blue, Sr \--{} green, Ru \--{} yellow, O \--{} red. \textbf{a)}: \sro\ surface structure with the ($1\times1$) unit cell \cite{Stoeger2014surf}. \textbf{b)-c)}: experimental and simulated STM, respectively. STM parameters: \stmpar{+0.05}{0.15}{78}. \textbf{d)}: \cro\ surface with the (1$\times$1) unit cell. The dashed lines indicate the characteristic wide and narrow channels caused by the tilting of the octahedra. \textbf{e)-f)}: experimental and simulated STM, respectively. STM parameters: \stmpar{+0.8}{0.1}{78}.} 
\label{fig:perovskites-surfaces}
\end{center}
\end{figure}

Up to now, surface-science studies have mostly focused on  two specific materials, \sro\ \cite{Stoeger2014co,Stoeger2014surf,Halwidl2015} and \cro\ \cite{Halwidl2017,Halwidl2018}, with a similar electronic and geometric structure. The repeating unit consists of two vertically stacked perovskite ARuO$_{3}$ (A = Sr, Ca) cells with RuO$_{6}$ octahedra separated from the next unit by a rock-salt-like AO type interface. At this interface, the layered structure is easily cleaved \cite{Stoeger2014surf}, leading to large, pristine AO terminated ($001$) surfaces. The bulk structures of these two particular AO-terminated materials show small, but significant differences in the arrangement of the RuO$_{6}$ octahedra, which leads to distinct differences in their surface structures. While the RuO$_{6}$ octahedra are only rotated around the [$001$] axis in \sro\ \cite{hu2010}, they are both rotated and tilted in \cro\ \cite{yoshida2005}. Therefore, the surface of \srosurf\ is identical to that of a binary SrO oxide, albeit at a larger lattice constant, while the additional tilting leads to alternating wide/narrow channels with decreased/increased oxygen density along the [$010$] direction on a \crosurf\ surface, see Fig.~\ref{fig:perovskites-surfaces}. 

This distinct difference in the surface structure strongly influences the adsorption behavior of various small molecules including H$_{2}$O and \oo. As shown recently, water adsorbs in a dissociated configuration on both \sro\ \cite{Halwidl2015} and \cro\ \cite{Halwidl2017} at low coverage. The split-off proton forms a surface hydroxyl with an apical oxygen of a RuO$_{6}$ octahedron, holding the remaining \oha\ fragment adsorbed at an A--A cation bridge in place via formation of a hydrogen bond. However, on \sro, this \oha\ can hop around the surface hydroxyl with a low activation energy of 187\,meV \cite{Halwidl2015}, while on \cro\ the channels caused by the tilting of the octahedra prevent a hopping of the \oha\ fragment \cite{Halwidl2017}. At higher water coverages, rows of dissociated H$_{2}$O are formed on \srosurf\ along the [$100$] and [$010$] directions, which finally connect to cages and capture molecularly adsorbed water \cite{Halwidl2015}. In  contrast, on \crosurf\ the higher reactivity of the surface results in only dissociative adsorption of water, and the resulting hydroxyl structures extend only along the preferred surface channels. Molecularly adsorbed water is found only in a second layer \cite{Halwidl2017}. 

The adsorption behavior of oxygen also reflects the structural differences of the (001) surfaces of \sro\ and \cro. In both cases, \oo\ adsorbs as a superoxo \oom\ species by electron transfer from the valence band of the substrate \cite{Halwidl2018,Mayr2018}. At low coverages, the adsorbed \oom\ molecule sits at an A--A cation bridge site in a slightly tilted geometry. On \cro, the channel structure of the surface caused by the tilting of the octahedra results in a stark preference of the \oom\ to adsorb at wider channels where the octahedra are tilted away from each other, while the higher symmetry of the \sro\ surface does not  give preference to a specific A--A cation bridge site. This pattern persists at high coverage, where on \cro\ a zig-zag pattern is formed that follows the wider channels \cite{Halwidl2018}. On \sro\ a similar zig-zag pattern of \oom\ adsorbed at Sr--Sr bridge sites is predicted, but without a directional preference \cite{Mayr2018}.

The adsorption of CO molecules has so far only been studied on the \srosurf\ surface \cite{Stoeger2014co}. There, the CO adsorbs weakly above an apical surface oxygen  at liquid nitrogen temperature. After heating the sample to 100\,K, the CO is activated and forms a metal carboxylate. This occurs by breaking the surface Ru--O bond and forming a bent COO with C bound to the Ru atom beneath. Due to the high symmetry of the surface, the surface carboxylate can rotate by 90\degree\ with a calculated barrier of 0.44\,eV, which can be overcome by scanning with the STM tip at +2.4\,V sample bias voltage \cite{Stoeger2014co}. 

In this work we show that \crosurf\ shows a similar behavior as \srosurf\ with regard to CO adsorption, albeit with differences attributed to the specific surface symmetry. CO adsorbs weakly at 78\,K, but activation either by annealing or manipulation with the STM tip causes the formation of the same Ru--COO carboxylate complex as on \srosurf. However, the surface carboxylate is always located at the edge of a wide channel and one rotational direction is preferred due to the reduced surface symmetry.

\section{Methods}

\subsection{Experiment}

We have used high-quality calcium ruthenate single crystals grown by the floating zone technique using a mirror-focused furnace \cite{Bao2008}. Before insertion into the UHV, the samples were fixed on stainless-steel sample plates with conducting silver epoxy glue (EPO-TEK H21D, Epoxy Technology Inc.), and a metal stud was glued on top with another epoxy adhesive (EPO-TEK H77, Epoxy Technology Inc.). The crystals were cleaved in UHV at 110\,K by removing the metal stub with a wobble stick.
STM measurements were carried out in a UHV system consisting of a preparation chamber and an STM chamber with base pressures of $2\times10^{-11}$ and $6\times10^{-12}$\,mbar, respectively. A low-temperature STM (commercial Omicron LT-STM) was operated at 78\,K in constant-current mode using an electrochemically etched W-tip. For all STM measurements the bias voltage was applied to the sample; positive or negative bias voltages result in STM images of the unoccupied or occupied states, respectively. All STM images shown were corrected for distortions as described elsewhere \cite{Choi2014}. CO was dosed to the STM head at a temperature of 78\,K, with a leak valve in direct line of sight to the sample. Since CO not reaching the sample is efficiently pumped by the walls of the cryostat, the gas doses measured with a Bayard Alpert gauge are very inaccurate.

\subsection{Computational Methods}
The DFT calculations were performed with the Vienna \textit{Ab-Initio} Simulation Package (VASP) using the projector-augmented plane wave method \cite{Kresse1999}. The surface structure was modeled using a double-layer slab of \cro, separated by 21\,\AA\ of vacuum. The adsorption of CO molecules was studied using a (2$\times$2) and a (3$\times$3) model cell, yielding a coverage of 1/8 and 1/18 monolayer (ML), defined by the number of octahedra in the surface unit cell, respectively. 
The electronic interactions were described using the van-der-Waals corrected \opt\ \cite{Klimes2010,Klimes2011} exchange-correlation functional. For comparison with the results for \sro, selected configurations were also computed with the standard PBE functional \cite{pbe1996}. The Brillouin zone was sampled on a $3\times3\times1$ $\Gamma$-centered Monkhorst-Pack $\vec{k}$-point grid and the kinetic energy cutoff was set to 400\,eV. STM simulations were performed within the framework of the Tersoff-Hamann approximation \cite{Tersoff1983}.

\section{Results}

\subsection{Initial Adsorption}
\begin{figure}[htbp]
	\centering
	\includegraphics[width=0.95\textwidth]{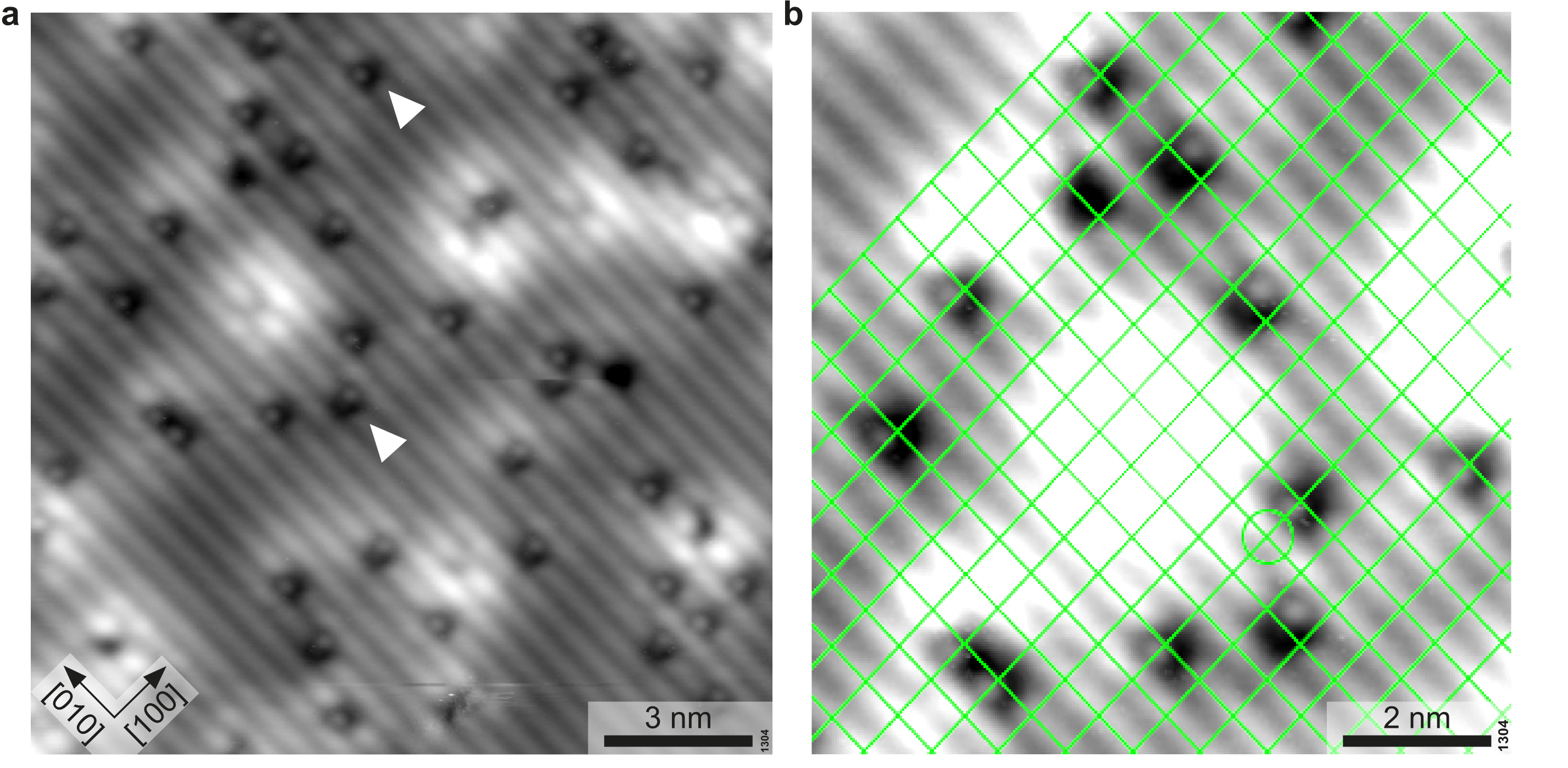}
	\protect\caption{\textbf{Initial adsorption of CO.} \textbf{a)}~STM image after dosing 0.002\,L CO. The molecules appear as dark depressions (marked by \textit{triangles}) on the bright substrate channels.  \textbf{b)}~An overlay of lattice unit cells shows the registry of the adsorption sites. Inequivalent sites are shifted by a half-integer number of lattice constants along [$010$] relatively to each other. STM parameters:  \stmpar{+0.4}{0.1}{78}.}\label{fig:exp-coinitial}
\end{figure}

As a first step, a small, nominal exposure of 0.002 Langmuir (L) CO was dosed onto the freshly cleaved \crosurf\ surface at 78\,K. The CO molecules form dark spots on top of the bright channel lines in the STM images taken at +0.4\,V bias voltage (see Fig.~\ref{fig:exp-coinitial}a). At this low coverage, the CO molecules appear as singular entities, and only a few pairs can be found, recognizable as larger dark spots elongated along the [$010$] direction. Additionally, a small bright dot appears at the adsorption site, displaced from the center of the bright channel in either [$100$] or [$\bar{1}00$] direction. The faint dark spots with a bright halo are attributed to point defects (impurities) originating from crystal growth.  It should be emphasized that the contrast in STM images is very sensitive to the tunneling conditions, and varies substantially with tip termination.  This is true for both \sro\ and \cro, likely due to their strongly-correlated electronic structure. 

The overlay of lattice unit cells, shown in Fig.~\ref{fig:exp-coinitial}b, reveals that the CO molecules occupy two distinct adsorption sites in the unit cell. Molecules with the small bright dot oriented towards [$100$]/[$\bar{1}00$] are separated by full/half integer numbers of lattice constants along [$010$]. 

\begin{figure}[htbp]
	\centering
	\includegraphics[width=0.95\textwidth]{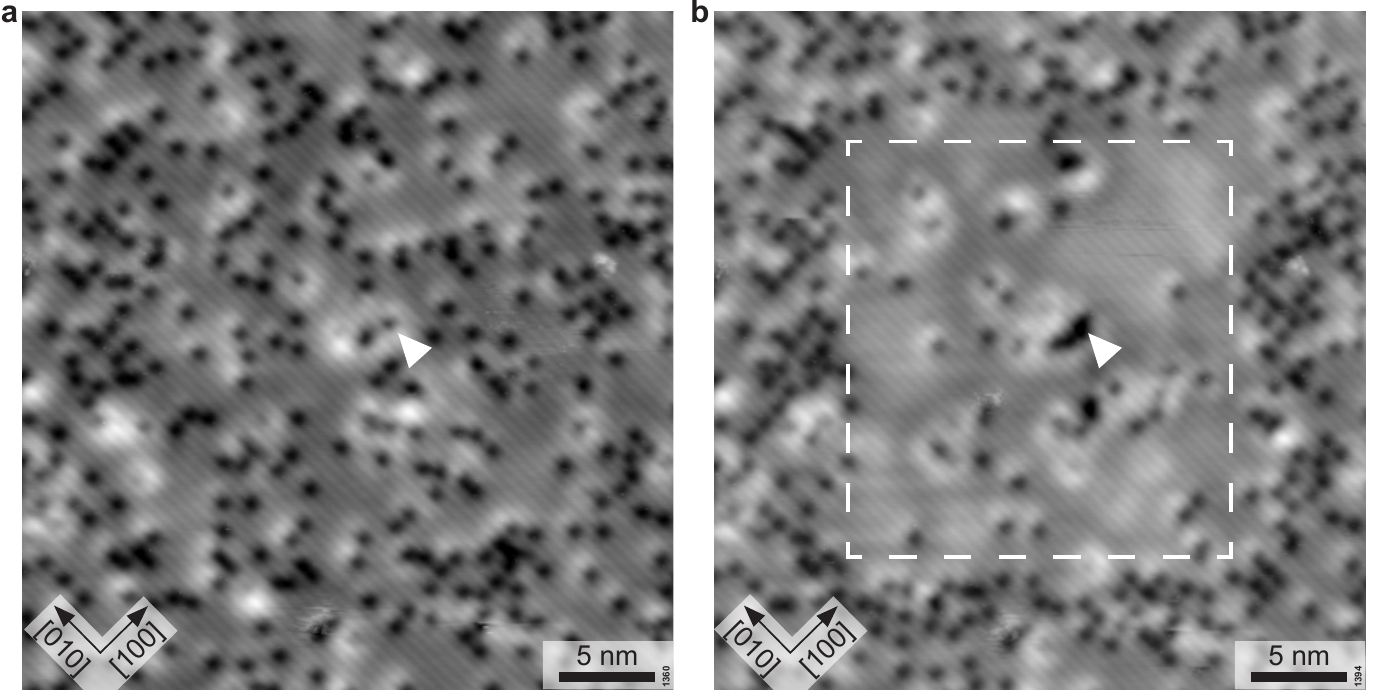}
	\protect\caption{\textbf{CO manipulation by the STM tip.} \textbf{a)}~STM image of a sample with a cumulative CO dose of 0.004\,L. The \textit{triangle} marks a defect. \textbf{b)}~Same area as in panel \textbf{a}. The area marked by the \textit{dashed rectangle} was scanned 32 times with $V_{\mathrm{sample}}$ = +1.0\,V.
		STM parameters: \stmpar{+0.5}{0.1}{78}.}\label{fig:exp-coprecmove}
\end{figure}

At an increased sample bias voltage of +0.7\,V the adsorbed CO starts to be influenced by the tip. Consecutive STM images show that the dark spots are moved mostly along the [$010$] direction for a few lattice constants. Fig.~\ref{fig:exp-coprecmove} shows a comparison of a sample exposed to 0.004\,L CO before and after scanning the rectangular area at the center 32 times with a sample bias voltage of +1.0\,V ($I_{t}=0.1$\,nA). As a result, almost all molecules are moved to the edge of the scanned area with a few molecules clustering next to a surface defect, indicated by a white triangle in Fig.~\ref{fig:exp-coprecmove}b.

\subsection{Chemisorption}
\begin{figure}[htbp]
	\centering
	\includegraphics[width=0.95\textwidth]{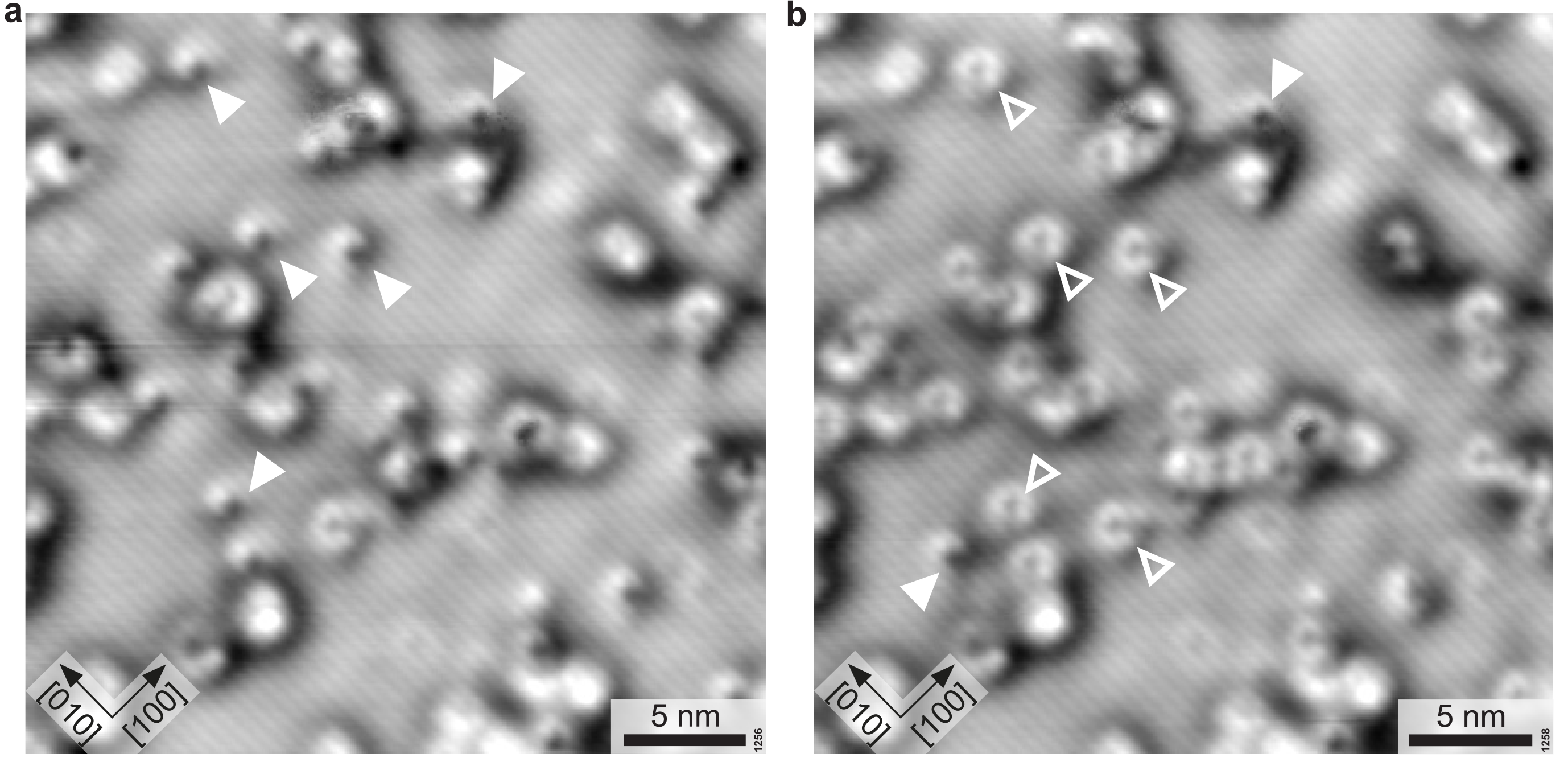}
	\protect\caption{\textbf{Transformation of the CO precursor into the chemisorbed species.} \textbf{a)}~The precursor species (\textit{full triangles}) appear as faint, dark spots with a bright secondary spot in the [$010$] direction. \textbf{b)}~Same area as in panel~\textbf{a}. After one scan with $V_{\mathrm{sample}}$ = +1.4\,V different species consisting of dark spots with bright, almost full halos have formed (\textit{open triangles}). They are attributed to chemisorbed molecules. Some precursor species (\textit{full triangles}) have remained or were moved to a new location. STM imaging parameters: \stmpar{+0.3}{0.1}{78}.}\label{fig:exp-cotransform}
\end{figure}

\begin{figure}[htbp]
	\centering
	\includegraphics[width=0.95\textwidth]{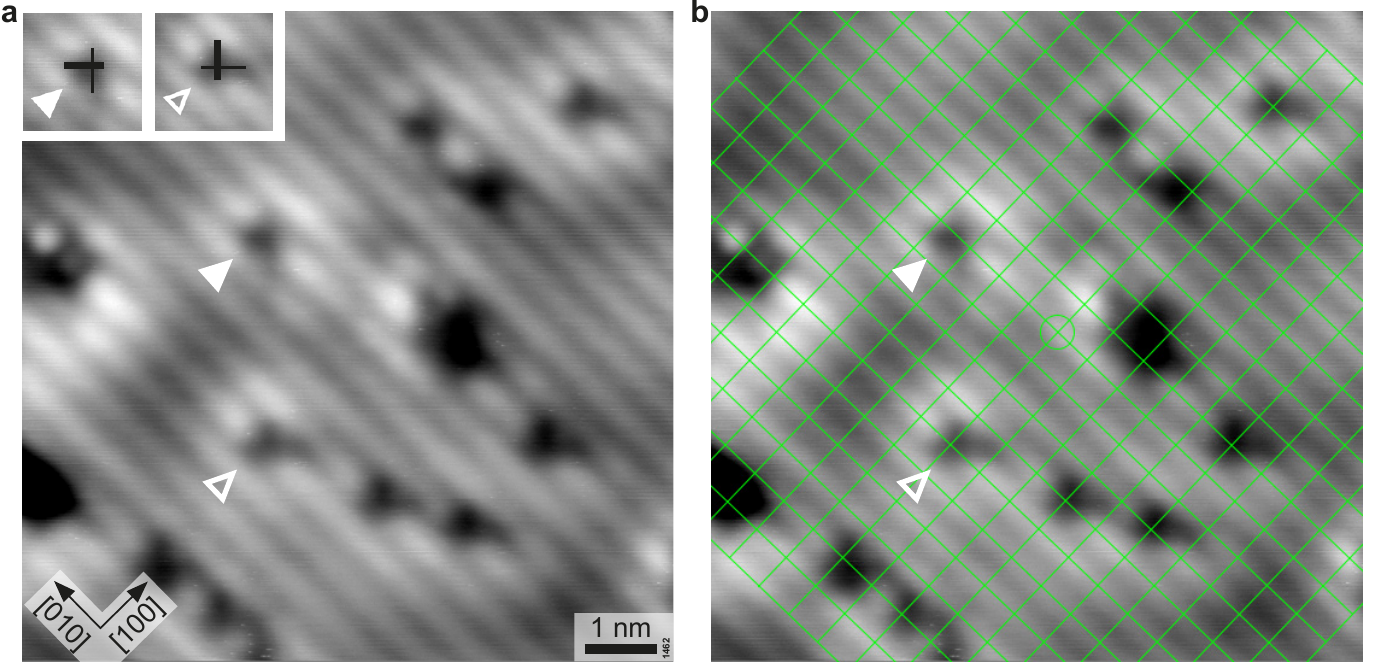}
	\protect\caption[STM imags of the chemisorbed species.]{\textbf{The chemisorbed species.} \textbf{a)}~Two similar orientations are observed. The insets show a sketch of the uneven, cross-like appearance. \textbf{b)}~Same area as in panel~\textbf{a}. The grid shows that the two orientations are shifted by a half\=/integer number of lattice constants along [$010$] relative to each other. STM parameters: \stmpar{+0.5}{0.1}{78}.}\label{fig:exp-cochemdetail}
\end{figure}

As on \srosurf\ \cite{Stoeger2014co}, the weakly bound CO can be transformed into a strongly bound species either by the STM tip or by annealing. In accordance with Ref. \cite{Stoeger2014co} the weakly bound species  is referred to as the precursor state. To transform the CO precursor to the strongly bound species on \crosurf\, an STM bias voltage of at least +1.4\,V is required. Alternatively, annealing at room temperature for 30\,min also resulted in the transformation. Fig.~\ref{fig:exp-cotransform} shows the tip-induced change. First, the precursor-covered sample was scanned at a  sample bias voltage of +0.3\,V where the CO molecules appeared as faint, dark spots located on the bright surface channels with a 
bright secondary spot in the [$010$] direction. After scanning with +1.4\,V, dark, uneven crosses with almost circular, bright halos have formed at the sites that were previously occupied by the precursor species, see Fig.~\ref{fig:exp-cotransform}b and Fig.~\ref{fig:exp-cochemdetail}a.

The chemisorbed species are better seen at higher resolution, as shown in Fig.~\ref{fig:exp-cochemdetail}. The uneven, dark crosses consist of two perpendicular, short lines, see the insets in Fig.~\ref{fig:exp-cochemdetail}a. Two distinct sites and orientations of the crosses are observed, shown by the overlaid grid of unit cells in Fig.~\ref{fig:exp-cochemdetail}b. As for the precursor species, identical features are found for integer shifts along [$010$], while the other orientation is found for half-integer shifts. Manipulation of the chemisorbed species requires higher sample bias voltages above +2.5\,V for CO movement. Above +3.0\,V sample bias voltage the CO is desorbed. 

\subsection{DFT}
\begin{figure}[tbhp]
	\centering
	\includegraphics[width=0.95\textwidth]{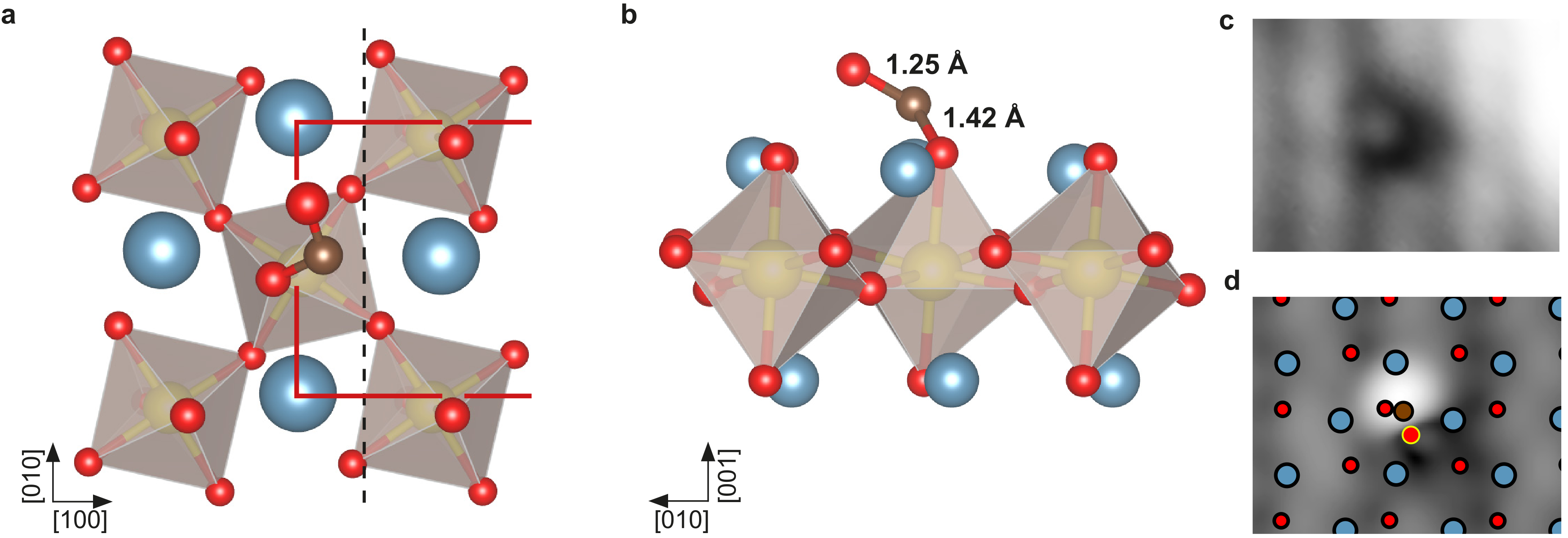}
	\protect\caption[DFT (2$\times$2) model of the CO precursor configuration]{\textbf{DFT (2$\times$2) model of the precursor configuration.} Ca \--{} blue, Ru \--{} yellow, O \--{} red, C \--{} brown. \textbf{a)}~Top view. The CO forms a C--\osurf\ bond that is tilted mainly towards [$100$]. The O of the CO is tilted towards [$010$]. The \textit{red line} and the \textit{dashed line} indicate the unit cell and the glide plane of the pristine surface, respectively. \textbf{b)}~Side view. \textbf{c)}~STM of the precursor, magnified from Fig.~\ref{fig:exp-coinitial}a. \textbf{d)}~STM simulation of the corresponding PBE (3$\times$3) model cell at +0.4\,V sample bias voltage. Yellow circles indicate the oxygen of the adsorbed CO.}\label{fig:dft-codftprec}
\end{figure}

DFT calculations show that the intact molecule adsorbs on top of an apical \osurf\ atom of a RuO$_{6}$ octahedron with a C--\osurf\ distance of 1.42\,\AA, right at the edge of a bright channel. The predicted binding energy of the CO is $-0.85$\,eV. As shown in Fig.~\ref{fig:dft-codftprec}, the molecule does not adsorb in a perfectly upright position. The C--\osurf\ bond is tilted by 37\degree\ towards the [$110$] direction, opposite to the tilt of the RuO$_{6}$ octahedron, and the O of the CO is tilted by 29\degree\ towards the [$010$] direction. The C--O distance of the adsorbate is elongated to 1.25\,\AA\ compared to the calculated gas-phase value of 1.14\,\AA. The apical \osurf\ is pulled slightly upwards toward the adsorbate, leading to an increase of the Ru--\osurf\ bond length by 9\%, to a value of 2.14\,\AA.  Tersoff-Hamann STM simulations (see Fig.~\ref{fig:dft-codftprec}d) show a pattern where the oxygen atom of the adsorbed CO induces a small dark spot along the bright channel line which the CO molecule is rotated towards, slightly shifted to one side. The carbon atom is indicated by an additional bright spot which is slightly shifted to the other side of the center line of the bright channel.

\begin{figure}[tbhp]
	\centering
	\includegraphics[width=0.95\textwidth]{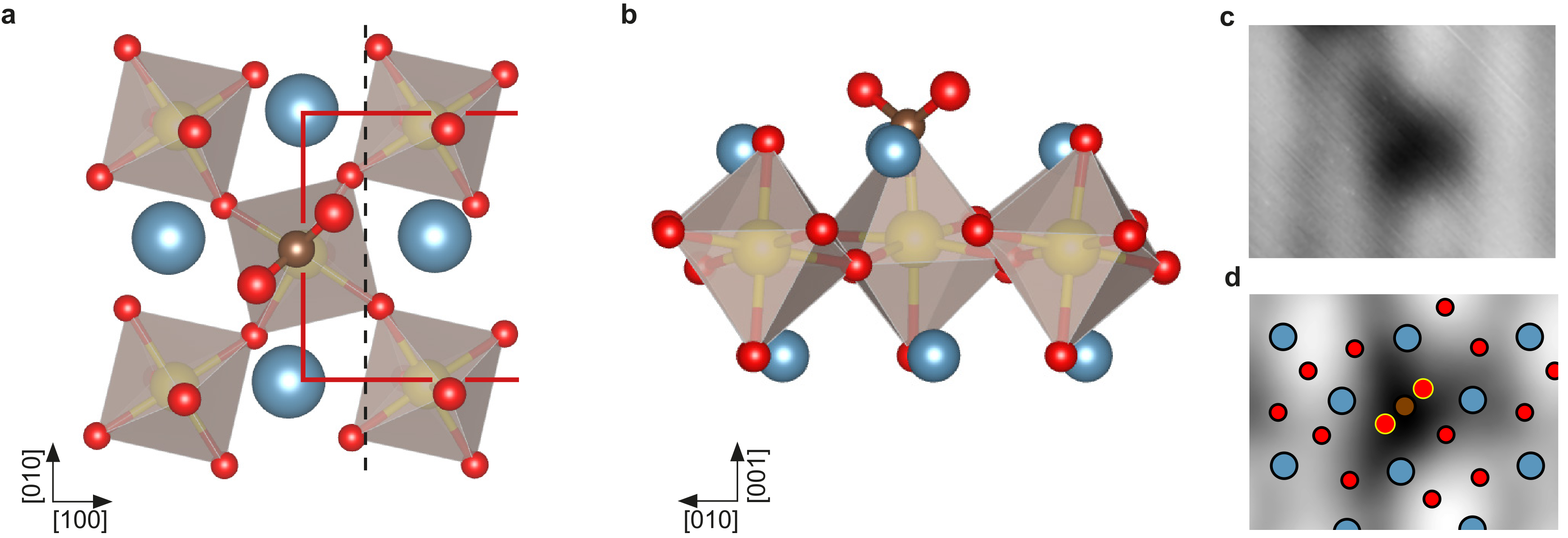}
	\protect\caption[DFT (2$\times$2) model of the chemisorbed Ru-COO configuration]{\textbf{DFT (2$\times$2) model of the chemisorbed Ru--COO configuration.} Ca \--{} blue, Ru \--{} yellow, O \--{} red, C \--{} brown. \textbf{a)}~Top view. The oxygen atoms of the Ru--COO are oriented along [110], thereby maximizing the distance to the neighbouring apical O; a configuration that is rotated by 90\degree\ is disfavoured by 70\,meV. The \textit{red line} and the \textit{black dashed line} indicate the unit cell and the glide plane of the pristine surface, respectively. \textbf{b)}~Side view. \textbf{c)}~Experimental STM of the carboxylate, magnified from Fig.~\ref{fig:exp-cochemdetail}a. \textbf{d)}~Simulated STM of a corresponding PBE (3$\times$3) model cell at +0.5\,V sample bias voltage. Yellow circles indicate the carboxylate oxygen.}\label{fig:dft-codftchem}
\end{figure}

The chemisorbed state of CO adsorbed on \srosurf\ \cite{Stoeger2014co} can be explained by replacement of the apical oxygen atom by the carbon atom of the adsorbate, resulting in the formation of a strongly-bound carboxylate. On the \crosurf\ surface, DFT predicts a similar mechanism. As shown in Fig.~\ref{fig:dft-codftchem}, the resulting carboxylate is located almost at the center of the surrounding Ca atoms and shows an adsorption energy of $-2.04$\,eV with respect to gas-phase CO. The O--C--O angle is 120\degree. The two oxygen atoms of the carboxylate are aligned with the [$110$] direction and point towards nearby Ca--Ca bridge sites. The O--Ca distances range from 2.40\,\AA\ to 2.55\,\AA. The Ru--C bond length is 2.04\,\AA\ and the Ru is lifted by 0.11\,\AA\ in [$001$] direction towards the surface compared to the pristine RuO$_{6}$ octahedron. Performing a rotation of the carboxylate by 90\degree\ reduces the stability by 70\,meV. This can be understood by inspecting Fig.~\ref{fig:dft-codftchem}a, which shows that for a 90\degree\ rotated carboxylate one oxygen is too close to the apical oxygen of the neighboring (top-left in Fig.~\ref{fig:dft-codftchem}a) RuO$_6$ octahedron tilted towards the carboxylate. DFT also predicts a higher barrier for rotation of $E_{\mathrm{barr}}= 0.56$\,eV than on \srosurf\ ($E_{\mathrm{barr}}= 0.44$\,eV). The STM simulations agree well with the experiment, see Fig.~\ref{fig:dft-codftchem}d. The carboxylate is shown as an uneven, large dark spot, slightly shifted from the bright channel line. 

\section{Discussion}
Both experimental and DFT results suggest a similar adsorption mechanism for the adsorption of CO on the (001) surface of \cro\  as for \sro. While the reported value for the adsorption energy of $-0.66$\,eV on \srosurf\ for the precursor species is significantly lower than the $-0.85$\,eV predicted in the current calculations, this can be attributed 
mainly to the different functional (including van-der-Waals corrections) in the present study. Using PBE like for \sro, a calculation for a larger ($3\times3$) model cell yields an even lower adsorption energy of $-0.59$\,eV. In both cases the bond lengths and the \osurf--C--O angle of the precursor are very similar, but the Ru--\osurf--C angle of 111\degree\ is significantly smaller than on \sro, where it is $\approx$143\degree. 

On \sro, St{\"o}ger et al.\ \cite{Stoeger2014co} have observed STM-induced desorption of the CO precursor at sample bias voltages of 0.3--0.4\,V, though as a very rare event. On \cro, we have never observed desorption of the precursor by STM; bias voltages of 1\,V or more (0.7\,V in rare cases) rather lead to mobility of the CO precursor on the surface. On \sro, in this voltage range the precursor transforms already to the carboxylate.

The chemisorbed mode is less strongly bound on \crosurf\ than on \srosurf; here the van-der-Waals corrected functional predicts an adsorption energy of $-2.04$\,eV (\sro: $-2.17$\,eV, PBE). Again, the PBE calculation on the ($3\times3$) model cell yielded an even smaller binding energy of $-1.91$\,eV, indicating a reduced reactivity of the \crosurf\ compared to the \srosurf\ surface. 

Comparing the functionals for \cro\, one clearly sees that van-der-Waals effects are more important for the precursor than for the chemically bound carboxylate. For the same functional (PBE)  the differences between \sro\, and \cro\, are more distinct for the carboxylate, which is not too surprising since the precursor barely influences the substrate geometry, while the carboxylate strongly affects the geometry (tilting) of the involved  RuO$_6$ octahedron.

While the bond lengths and angles of the carboxylate are virtually identical on both surfaces, it should be noted that on \crosurf\ the whole  RuCO$_{5}$ octahedron is not tilted, but that its $c$-axis is kinked by 9\degree\ at the RuO$_{4}$ basal plane. On \srosurf, a kinked O--Ru--C axis distortion does not occur, since the octahedra are not tilted in the first place and the apical position of the C is right above the Ru atom. It is also notable that both the precursor and the carboxylate adsorb above a surface O or on top of a Ru, respectively, in contrast to the adsorption of other molecules such as \hho\ and \oo, which adsorb at O--O bridge sites \cite{Halwidl2015,Halwidl2017,Halwidl2018}. 
On \srosurf, the tip-induced desorption of the chemisorbed mode at a sample bias voltage of +2.7\,eV was explained by electron capture into the lowest antibonding O--CO orbital at +2.4\,eV. Here, DFT predicts the same O--CO orbital slightly higher at +2.7\,eV, in agreement with the higher sample bias voltage of +3.0\,V needed to remove the carboxylate species on \crosurf. Due to the lower symmetry of the \cro\ unit cell, no tip-induced rotation of the COO carboxylate was observed. 

\section{Conclusions}
To conclude, we report a high reactivity of the \crosurf\ surface with regard to the adsorption of CO. The adsorption mechanism for both the precursor and the carboxylate is similar to the highly symmetric \srosurf\ surface, albeit adsorption energies are somewhat lower on \crosurf.
In contrast to other small molecules such as \hho\ and \oo, which prefer adsorption at O--O bridge sites, the CO molecules adsorb either at apical O (precursor) or above Ru (carboxylate) on \crosurf.

\section*{Acknowledgments}
This work was supported by the Austrian Science Fund (FWF project F45 ``FOXSI'') and the Vienna Scientific Cluster (VSC). The structural models were created with the program VESTA \cite{Momma:db5098}.


\begin{thebibliography}{10}

\bibitem{Mao2000}
Z.Q.~Mao, Y.~Maeno, H.~Fukazawa, 
{Crystal growth of Sr$_2$RuO$_4$}, 
Materials Research Bulletin 35 (2000) 1813--1824. 
%doi:10.1016/S0025-5408(00)00378-0.)

\bibitem{Stoeger2014surf}
B.~St{\"{o}}ger, M.~Hieckel, F.~Mittendorfer, Z.~Wang, M.~Schmid, G.~S. Parkinson, D.~Fobes, J.~Peng, J.~E. Ortmann, A.~Limbeck, Z.~Mao, J.~Redinger, U.~Diebold, 
{Point defects at  cleaved Sr$_{n+1}$Ru$_{n}$O$_{3n+1}$ surfaces}, 
Phys. Rev. B 90 (2014) 165438.

\bibitem{Stoeger2014co}
B.~St\"oger, M.~Hieckel, F.~Mittendorfer, Z.~Wang, D.~Fobes, J.~Peng, Z.~Mao, M.~Schmid, J.~Redinger, U.~Diebold, 
{High chemical activity of a perovskite surface: reaction of CO with Sr$_{3}$Ru$_{2}$O$_{7}$}, 
Phys. Rev. Lett. 113 (2014) 116101. %doi:10.1103/PhysRevLett.113.116101.

\bibitem{Halwidl2015}
D.~Halwidl, B.~St\"oger, W.~Mayr-Schm{\"{o}}lzer, J.~Pavelec, D.~Fobes, J.~Peng, Z.~Mao, G.~Parkinson, M.~Schmid, F.~Mittendorfer, J.~Redinger, U.~Diebold, 
{Adsorption of water at the SrO surface of ruthenates}, 
Nat. Mater. 15 (2015), 450--455.
 %http://doi.org/10.1038/nmat4512

\bibitem{Halwidl2017}
D.~Halwidl, W.~Mayr-Schm{\"{o}}lzer, D.~Fobes, J.~Peng, Z.~Mao, M.~Schmid, F.~Mittendorfer, J.~Redinger, U.~Diebold,
{Ordered hydroxyls  on Ca$_{3}$Ru$_{2}$O$_{7}$(001)}, 
Nat. Commun. 8 (2017) 23.

\bibitem{Halwidl2018}
D.~Halwidl, W.~Mayr-Schm{\"{o}}lzer, M.~Setvin, D.~Fobes, J.~Peng, Z.~Mao, M.~Schmid, F.~Mittendorfer, J.~Redinger, U.~Diebold,
{A full monolayer of superoxide: oxygen activation on the unmodified Ca$_3$Ru$_2$O$_7$(001) surface}, 
J. Mater. Chem. A 6 (2018) 5703--5713.

\bibitem{hu2010}
B.~Hu, G.~T.~McCandless, M.~Menard, V.~B.~Nascimento, J.~Y.~Chan, E.~W.~Plummer, R.~Jin,
{Surface and bulk structural properties of single-crystalline Sr$_{3}$Ru$_{2}$O$_{7}$}, 
Phys. Rev. B 81 (2010) 184104. %http://doi.org/10.1103/PhysRevB.81.184104

\bibitem{yoshida2005}
Y.~Yoshida, S.-I.~Ikeda, H.~Matsuhata, N.~Shirakawa, C.~H.~Lee, S.~Katano, 
{Crystal and magnetic structure of Ca$_{3}$Ru$_{2}$O$_{7}$}, 
Phys. Rev. B 72 (2005) 054412.% http://doi.org/10.1103/PhysRevB.72.054412

\bibitem{Mayr2018}
W.~Mayr-Schm{\"{o}}lzer, F.~Mittendorfer, J.~Redinger, 
{Adsorption of a superoxo O$_2^{\mbox{-}}$ species on the pure and Ca-doped Sr$_3$Ru$_2$O$_7$(001) surface}, 
submitted to Surface Science (2018)

\bibitem{Bao2008}
W.~Bao, Z.~Q.~Mao, Z.~Qu and J.~W.~Lynn, 
{Spin valve effect and magnetoresistivity in single crystalline Ca$_3$Ru$_2$O$_7$}, 
Phys. Rev. Lett. 100 (2008) 247203.

\bibitem{Choi2014}
J.~I.~J.~Choi, W.~Mayr-Schm\"olzer, F.~Mittendorfer, J.~Redinger, U.~Diebold and M.~Schmid, 
{The growth of ultra-thin zirconia films on Pd$_{3}$Zr(0001)}, 
J. Phys. Condens. Matter 26 (2014) 225003.

\bibitem{Kresse1999}
G.~Kresse, D.~Joubert, 
{From ultrasoft pseudopotentials to the projector augmented-wave method}, 
Phys. Rev. B 59 (1999) 1758--1775.

\bibitem{Klimes2010}
J.~Klime{\v{s}}, D.~R. Bowler, A.~Michaelides,
{Chemical accuracy for the  van der Waals density functional}, 
J. Phys. Condens. Matter 22 (2010) 022201.

\bibitem{Klimes2011}
J.~Klime{\v{s}}, D.~Bowler, A.~Michaelides,
{Van der Waals density functionals applied to solids}, 
Phys. Rev. B 83 (2011) 1--13.

\bibitem{pbe1996}
J.P.~Perdew, K.~Burke, M.~Ernzerhof,
{Generalized gradient approximation made simple}, 
Phys. Rev. Lett. 77 (1996) 3865--3868.

\bibitem{Tersoff1983}
J.~Tersoff, D.~Hamann, 
{Theory and application for the scanning tunneling microscope}, 
Phys. Rev. Lett. 50 (1983) 1998.

\bibitem{Momma:db5098}
K.~Momma, F.~Izumi, 
{{\it VESTA3} for three-dimensional visualization of crystal, volumetric and morphology data}, 
J. Appl. Cryst. 44 (2011) 1272--1276.

\end{thebibliography}
\end{document}